# Blockchain in Healthcare and Medicine: A Contemporary Research of Applications, Challenges, and Future Perspectives

A PREPRINT

**H. Sami Ullah**
Dept. of Computer Science
*University of Gujrat, Gujrat, Pakistan*
hamzasamiullah@gmail.com

**S. Aslam**
Dept. of Computer Science
*University of Gujrat, Gujrat, Pakistan*
samiaaslam479@gmail.com

## Abstract

Blockchain technology is one of the most contemporary and disruptive technologies in the world. It has gained considerable attention in numerous applications such as financial services, cybersecurity applications, Internet of Things (IoT), network data management. Now its range of applications is beyond the financial services as healthcare industry has also adopted blockchain technology in its various subdomains such as Electronic Health Records (EHR), medical supply chain management system, genomic market, neuroscience technology, clinical research, and pharmaceutical medicine. Blockchain is considered a secure and viable solution for storing and accessing patients medical records and the patients can diagnosed and treated with safe and secure data sharing. Blockchain technology will revolutionize the healthcare systems with personalized, authentic, and secure access to the clinical data of patients and that data can be used for further health improvements and clinical researches. In this paper, we conduct a contemporary research on existing applications and developments in healthcare industry with the use of blockchain technology. We also discuss some robust applications and various existing companies that are using blockchain solution for securing their data along with some current challenges and future perspectives.

## 1. Introduction

With the growth of population, the healthcare system and services are considered one of the top priorities in the modern society to provide better health facilities to the people. Data (medical records laboratory test reports, bill payments, clinical trials) generation rate in the healthcare industry is higher than any other industry. In the past, paper-based medical records were used as a repository of data and information about patients and were reviewed by various staff members for different purposes. These records were accessible by only one user at a time and the management of records in a systematic way was difficult. Digital transformation of healthcare records with electronic methods and techniques improves data management in the healthcare industry.

In the healthcare industry and clinical centers, security, privacy, and scalability of health records storage and sharing is highly imperative as the diagnosis and proper health decisions about the patient condition depend on their accurate data. The data shared by clinical practitioners for proper follow-ups of the patients must be transferred with privacy and data must be up-to-date according to the health conditions of the patients. Telemedicine and e-health services are also the most prominent domain of healthcare where data is transferred to the care professionals to diagnose and treat patients on remote locations using telecommunication technologies [1], [2]. Telemedicine is the striking evolution and has become an important part of healthcare infrastructure. Security, sensitivity, and privacy in these online patient monitoring can be a major challenge due to the sensitive data of patients. Thus, healthy and accurate communication between patients and physicians is directly linked with the accurate and secure data transactions [3]–[5].

Moreover, interoperability among the healthcare industry and various research organizations is the greatest challenge to be overcome for successful and secure data exchange. The constraints between a healthy collaboration of parties involved in data exchange are the variety like clinical records, various data-sharing agreements, the sensitivity of data, and some ethical policies and these aspects are imperative to be addressed while practical data exchange among different entities. Many kinds of research have been conducted in the past few years and the contemporary solutions to the healthcare industry are proposed with latest technologies such as Artificial Intelligence [6], Internet of Things (IoT) [7], machine learning, deep learning, and computer vision for more efficient and robust healthcare infrastructure for the betterment of human beings.

Among all these technologies, blockchain technology has a remarkable impact to provide more safe and secure healthcare infrastructure including healthcare record-keeping service [8], [9], biomedical field, medical supply chain management [10], telemedicine [5], genome data [11], and e-health data sharing services. Blockchain is the buzzword of the year and this technology slowly emerges from banking to the supply chain logistics [12]. In healthcare infrastructure, blockchain provides massive opportunities to lead the digital transformations of medical records, pharmaceuticals supply chains, and smart contracts for payment distribution and plenty of other methods to leverage this technology in the healthcare industry. As one of the most important features of a hospital is the health record of patients and electronic health records are the backbone of every modern healthcare center. But the medical records of patients grow longer with each visit to the doctor and become more complex. Every doctor and hospital has different methods to store these records, but this is not easy for healthcare providers to obtain them.

There are some companies such as the Patientory, MEDIBLOC, and MEDICALCHAIN aimed to solve this problem. The goal of these companies to give patients authority over the entire medical history and to provide one-stop access to it for the patients and physicians while inherently bringing data security to the field as well. Medical supply chains are also an important part of the healthcare system as the pharma industry has higher standards for product safety, security, and stability. Supply chain management with blockchain can be monitored securely and transparently by reducing time delays and human mistakes. It can also be used to monitor costs, labor, and even ways to emission of every point in the supply chain. It can be used to verify the authenticity of the products by tracking them from their origin combating the market that costs 2 hundred billion dollars in losses to the market annually. Companies like CHRONICLED,

BLOCKPHARMA, and modum are working towards more efficient logistics solutions. The genomics market is also using blockchain technology like the companies EncrypGen (A genomic blockchain network) and NEBULA GENOMICS are building blockchain platforms where people can share their genomic data safely and securely. These companies aimed to use blockchain to enhance data protection and enable buyers to efficiently acquire genomic data. These companies are just a few of these dozens of startups that aimed to use blockchain to disrupt the healthcare industry.

Blockchain is a distributed ledger with a growing list of records in the form of a chain of blocks that are linked with each other using cryptographic algorithms. Each block in the network is connected via the cryptographic hash value of its previous block and every block has a copy of transaction data. Blockchain follows a peer-to-peer network and has a potential edge to revolutionize the digital world with its key elements such as decentralization, transparency, anonymity, immutability, autonomy, and open-source access. These key elements are depicted in the following Figure 1:

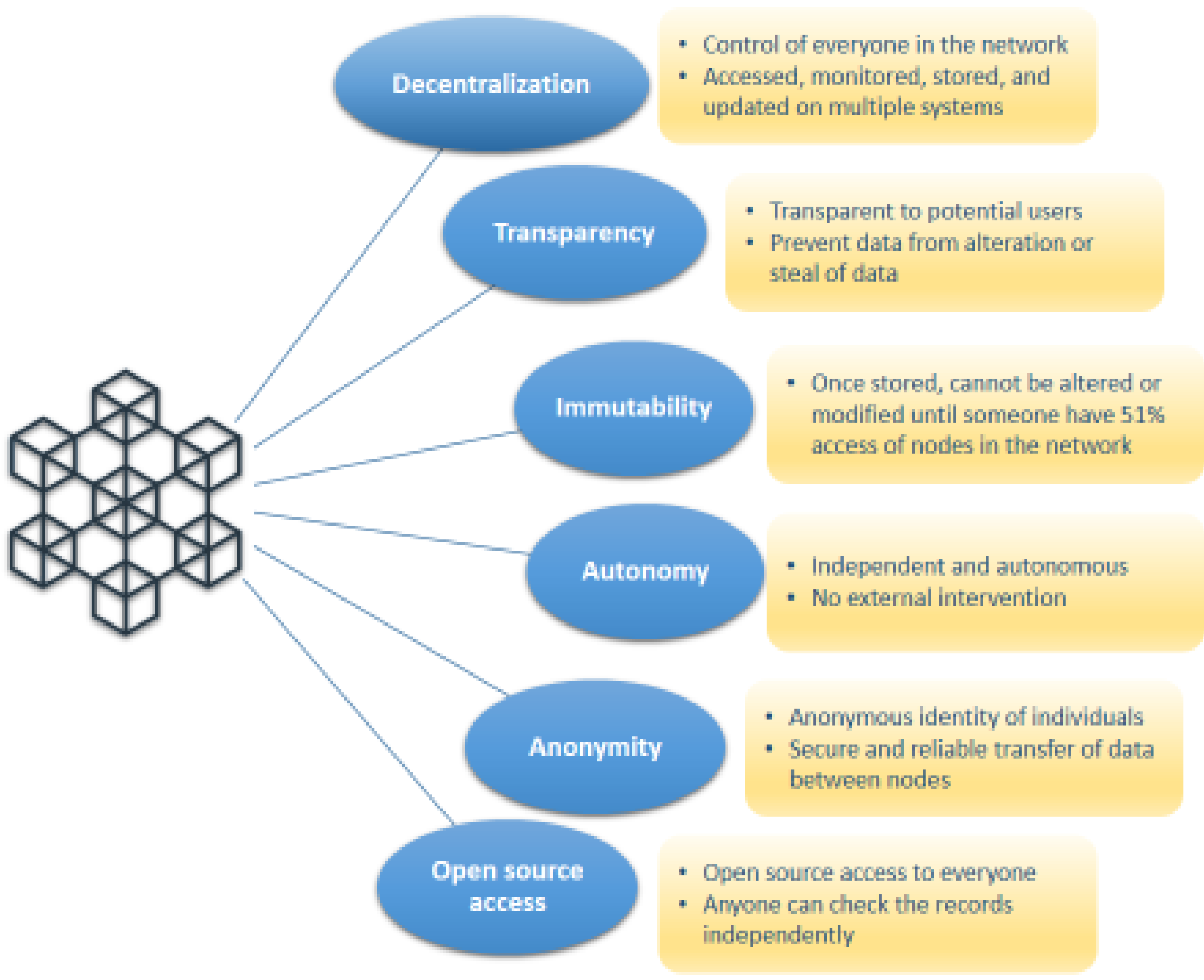

**Figure 1: key Elements of Blockchain**

All the transactions in the blockchain network are approved with the consensus process. Consensus in the network acts as an agreement among all nodes to validate the new blocks in the chain. However, Blockchain is a system where multiple connected computers maintain a secure and updated distributed ledger without any central authority. The P2P network of blockchain transactions is depicted in Figure 2 with a flow diagram:

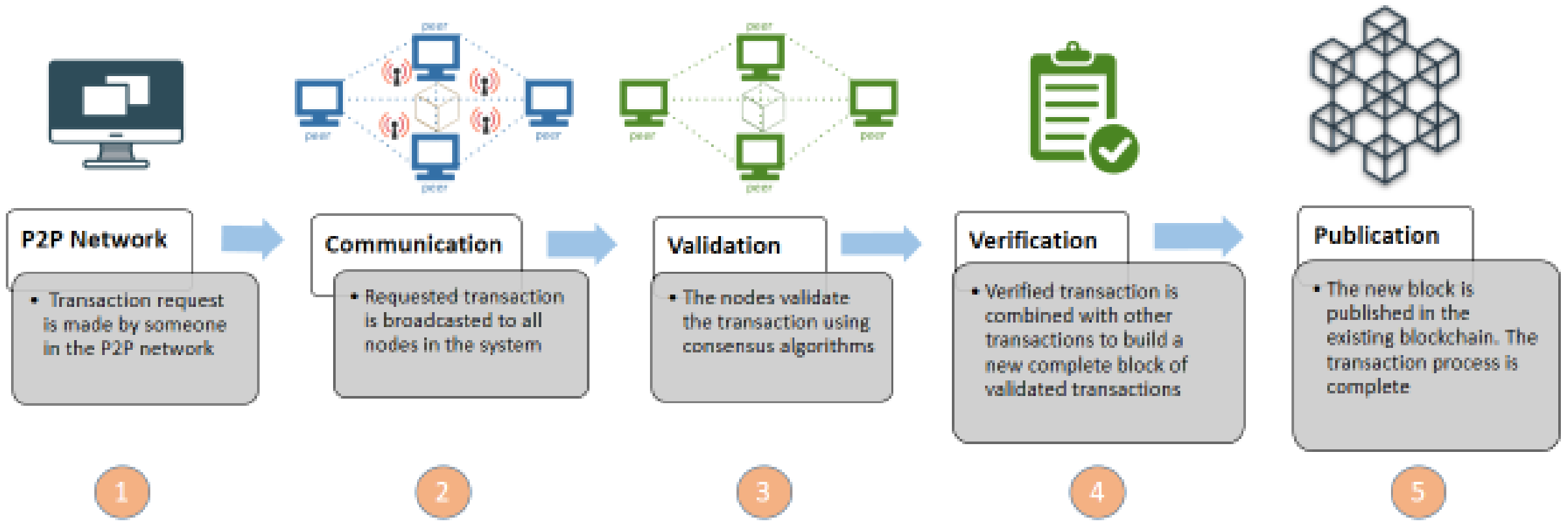

**Figure 2: Transaction Flow of Blockchain's P2P Network**

Blockchain has the potential to secure healthcare and medical data with the salient features of security and privacy over the distributed network. The clinical and medical data stored on a secure blockchain is beneficial for patients for their regular treatment and follow-ups with their physicians even in distant locations. The confidential data of patients become secure for their privacy and reliable for physicians for further improvements in healthcare applications and diagnosis.

In this paper, we will conduct contemporary research on the existing applications of blockchain in healthcare and medicine. The rest of the paper is organized as follows: Section 2 describes the related work done in healthcare industry with the integration of blockchain technology. Section 3 describes various applications of blockchain technology in healthcare and medicine industry. Section 4 gives an overview of companies using blockchain as a solution in terms of validating and maintaining their privacy and security via blockchain. Section 5 presents some current challenges faced in healthcare and medicine while using blockchain at a large scale. Section 6 highlights some future perspectives of blockchain technology in the healthcare industry. Section 7 concluded the paper followed by References section.

## 2. Related Work

The Healthcare industry is looking to adopt the blockchain technology in their healthcare applications for many years and is being researched in many previous studies. Private Blockchain is the best way to monitor and store the healthcare records with safety and security. A personalized healthcare method based on a private blockchain named Healthcare Data Gateway (HDG) was introduced in [13] where patients can access, monitor, and manage their records independently. In [14] a public blockchain with secure encryption methods was demonstrated for storing the health records of patients with security. In this approach, data was stored in encrypted codes and stored publically to the blockchain network where patients were enabled to access and monitor their records. A private blockchain based on Ethereum was proposed in [15] to implement safe and secure medical sensors and secure remote monitoring of patients in distant locations. This approach was used by practitioners to track the status of their patients with real-time monitoring to maintain a safe and up-to-date record of patients.

To manage and share a patient's medical records, another study [16] proposed a framework based on the integration of blockchain and cloud storage. This proposed scheme was used to achieve the safe and secure storage and exchange of personal medical data and information of patients. The suggested approach was used to give patients complete access and control to their records without the involvement of any third party. To evaluate the status of diseases in patients, a blockchain framework was proposed in [17] that used parallel execution and artificial healthcare systems to assess the overall condition, diagnosis, and treatment process of the patients. The therapeutic procedures were analyzed then through parallel executions and computational trials for making an appropriate decision about patient health conditions according to their disease. The suggested system was tested on real and artificial healthcare systems to test the accuracy of diagnosis and effectiveness of recommended treatment. In [18] a unique blockchain-oriented platform BloCHIE was developed for healthcare data and information exchange. The suggested platform implemented blockchain in various contexts to handle the requirements of personal data records and electronic healthcare records. To check the validity of the platform in terms of authenticity and privacy, they verified the platform with on-chain and off-chain verification systems.

Blockchain technology is encouraged to be adopted as a mechanism for sharing medical data among clinical specialists and healthcare providers with improved privacy, and security. A blockchain mechanism was proposed in [19] as an innovative architecture for the saving and maintaining the electronic health records of patients to protect their sensitive data and to address critical data security issues by implementing a blockchain software infrastructure throughout a hospital system. Blockchain technology has also proved its existence with the great potential in biomedical research, clinical data domains, and medical supply chain management. With the utilization of blockchain technology, clinical agreements, plans, and protocols can be stored on a blockchain network before commencing a clinical examination and the sensitive data related to clinical trials becomes up-to-date, secure, and publicly transparent. Smart contracts in the blockchain can also be deployed, replicated, and executed within various phases of clinical research to ensure transparency in the network. In [20] a blockchain-based telemonitoring framework was proposed for the diagnosis and treatment of cancer tumors for patients in remote locations. The suggested framework used smart contracts along with blockchain for ensuring the validity and security of the patient's data at specialized medical centers as well as in their homes. A blockchain-based method called DermoNet is suggested in [21] for the assistance of dermatology patients with online dermatological consultation mechanisms through teledermatology monitoring. ProActive Aging is a blockchain-based platform proposed in [22] to support the active living of aging people. Blockchain technology can be an ideal and well-suited choice for extensive medical treatment processes, such as chronic diseases, surgical operations, and aging. Moreover, pharmaceutical industries, drug manufacturers, and biomedical researchers are adopting blockchain to conduct advanced research at the genomic level in the healthcare industry.

## 3. Applications of Blockchain in HealthCare

Blockchain technology was originally implemented in banking and finance applications in the form of cryptocurrencies, but over time its potential has expended in various domains including healthcare and biomedical field [23]. The utilization of blockchain technology in the healthcare domain can be viewed in Electric Health Records (EHR), medical supply chain management, clinical research and development, genomic market, pharmaceutical medicine, neuroscience studies, biomedical development, and Telemedicine and E-Healthcare. Blockchain provides a stable and secure mechanism to store and share data in all sub-domains of the healthcare industry so that stored data can be used for different types of transactions and experiments by physicians and healthcare providers. Application of blockchain in the healthcare industry is discussed below in the light of some contemporary researches in the healthcare industry and depicted in Figure 3:

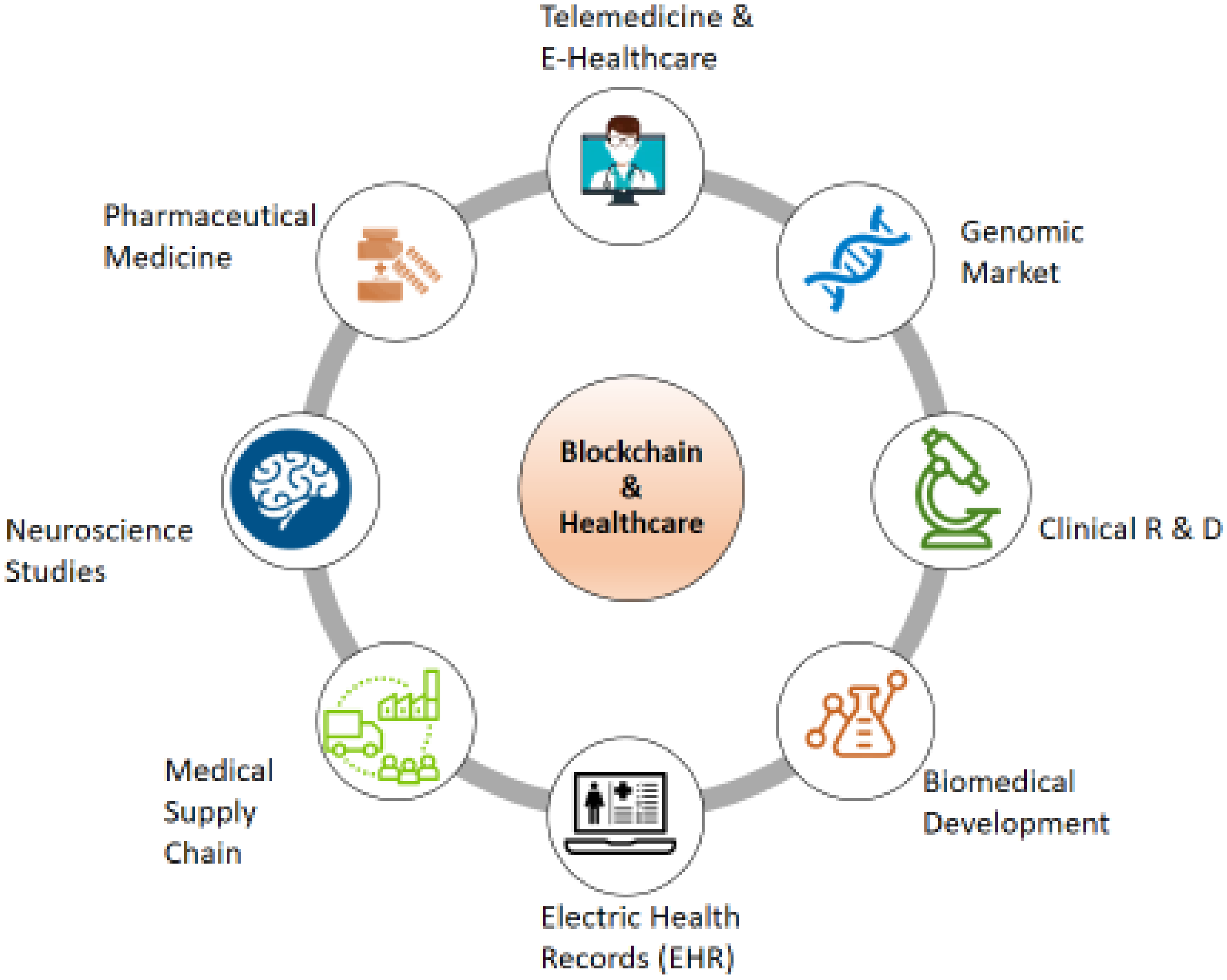

**Figure 3: Applications of Blockchain in Healthcare System**

### 3.1. Electronic Health Records (EHR)

The need for digitization in healthcare records is increased over the past decade as medical practitioners and healthcare providers need easy access to patient data for quick decision making. Electronic health records (EHR) are the digital version of data and these records are available at any time with enhanced security to authorized healthcare providers. EHR contains a patient's medical history, diagnosis reports, medication details, and treatment plans, laboratory and test results, etc. The most common application of blockchain in healthcare systems is in the electronic health records to make data more secure and reliable.

The limitation even in electronic health records before blockchain technology was the data of patients were scattered among various healthcare providers according to the situation of patients and the past data was not accessible even in EHR systems [24]–[26]. Blockchain is suggested by many researchers as a unique solution for storing patients' EHRs to keep their history and current information secure for a lifetime and can be retrieved at any time.

For providing an up-to-date and immutable record history of patients, a prototype “MedRec” was proposed in [27] that has different blockchain perks for managing authentication, integrity, security, confidentiality, and sharing of data records. This prototype works under a decentralized management system and ensures that the patients should have an immutable and easy-to-access history of their medical data that they can easily share with their various healthcare providers for treatments. Medical data sharing can become critical with some limitations while implementing records with EHR, such as loss of control over records, data provenance issues, and data trailing. To overcome these limitations, a blockchain-based platform named MeDShare was introduced in [28] for the secure and trustless exchange of data and electronic health records among different cloud service providers, healthcare physicians and research entities with improved data provenance and privacy.

For sharing of healthcare data with safety and security, a blockchain and MedRec-based approach was implemented where signcryption and attribute-based authentication methods were used for secure sharing of data in [29]. For the storage and processing of heterogeneous medical data, in another study [30] a blockchain-based framework is proposed for storing and managing heterogeneous electronic medical records with security and accuracy in a Cloud environment. Secure access to medical data by various entities is also imperative and is addressed in [31] where an Ethereum-based blockchain framework named Ancile with smart contracts was proposed for secure, and efficient access to medical records by various entities including patients, healthcare providers, and third parties. Another study [32] has proposed a blockchain-oriented authorization method called GAA-FQ (Granular Access Authorization supporting Flexible Queries) with granularity control.

## 3.2. Medical Supply Chain

The drug supply chain is one facet of the pharma landscape where blockchain can benefit from the unique characteristics and key principles of technology. The pharmaceutical supply chain is considered most prominently while deploying technology-driven solutions and use cases [33]. The counterfeit and fake global market is reaching up to $200 billion per year [34] and counterfeit medicines are a great troublesome as the global black market is providing such drugs to people without the ethical permissions. The risk of human life due to counterfeit drugs and medicines is increasing and cannot be underestimated. Counterfeiting is a significant problem within the pharmaceutical industry and must be considered an alternative solution to this problem. Blockchain technology has the potential to support supply chain transformation in the healthcare industry to reduce fraud with anti-counterfeiting systems and better management of quality in the manufacturing and distribution of pharmaceutical products and medicines.

Blockchain-based approaches offer the potential to track the whole lifecycle of drugs and medicines down to a single dose from making and distributing products like pharmaceuticals. Blockchain technology for secure digital marking of pharmaceutical products to provide a secure and reliable track of products throughout their lifecycle is already deployed by various companies such as Blockpharma, TIERION, CHRONICLED, Centers for Disease Control and Prevention (CDC).

## 3.3. Clinical Research and Development

Data privacy, integrity, sharing, record keeping, patient enrolling is a very sensitive feature in clinical trials and various issues of privacy and security may arise [35]. Blockchain provides viable solutions to these problems as being the next internet generation [36]. Many healthcare researchers are working on resolving the limitations and issues in clinical records with the help of blockchain technology [37], [38]. The healthcare industry will be more revolutionized with the applications and integration of blockchain, artificial intelligence (AI) and machine learning techniques. In [37], a permissioned Ethereum protocol with smart contract functionality [38], [39] is used for clinical data management in systems to address the issue of the patient enrollment. The Ethereum protocol is quicker in transactions as compared to bitcoin, and they proposed the use of Ethereum smart contracts for transparency and security of data management systems in clinical trials and researches. Therefore, patient enrolment using blockchain is one of the prominent existing applications of this technology in clinical research and trials. Another research was conducted [40] where a blockchain-based framework was implemented to track and store the informed consent of patients with security. These records will be secure, publicly verifiable, and unfalsifiable.

### 3.4. Genomic Market

Human genomic projects are generating a large volume of genomic data that is extensively used in the field of biotechnology and medical research. However, there is also a rapid increase in the throughput of gene sequencing technologies and scientists are enable to expedite the genome sequencing process to achieve significant cost benefits from it. It is estimated that around 15 % of people in the world's population will have their genome sequenced by 2025 and this generated data will be in zettabytes. However, no reliable and secure data resources are available to store a large volume of genomic and clinical data. So there is a need for such tools and technologies that can help in the processing and analysis of genomic data and can be easily accessible by physicians, scientists, pharmaceutical companies and other healthcare providers.

The blockchain technology has emerged as an appropriate and contemporary solution for storing and exchanging genomic data with safety and security. According to a report [41], the global blockchain in genomic market analysis grows at a CAGR of 66.42% during 2019-2029. In [42] to generate and access genomic data with privacy-preserving and decentralized methods that is blockchain is suggested with some highlighted recent challenges. In another study [43] current developments using blockchain technology to solve the genomic data issues are introduced and possible future implications are also highlighted. Various companies are using blockchain technology in genomic data generation and storage such as Nebula Genomic, EncrypGen, Doc.AI, Health Nexus, Luna Coin, zenome.io.

### 3.5. Pharmaceutical Medicine

The pharmaceutical and medicine sector is the largest growing and leading sector in the healthcare industry. The pharmaceutical sector introduces new and potential medicines and drugs with contemporary researches on various health conditions of patients to provide the appropriate medication in time. It also assists to ensure the safety and validity of medicines and drugs sold out to the end-users. The pharmaceutical sector also helps in the evaluation and processing [44] of safe drugs and medicines for quick recovery of patients.

In some usual cases, drugs and medicine companies cannot track their medical product timely, as sometimes the drugs have compromised with counterfeiters or fake drugs are invaded in the system. The production and distribution of fake medicines and drugs by counterfeit markets is a global risk to the health of people. Blockchain has the potential to evaluate, monitor, and ensure the safety and security in the production of potential drugs and medicines. In [45] a counterfeit medicine project based on Hyperledger blockchain was introduced for the inspection of the production of counterfeit drugs and medicines. For the effective delivery of reliable and authentic medical products to the patients, it is imperative to monitor, evaluate, and ensure the complete process of development and supply of pharmaceutical drugs and medicines. For the authenticity of delivered medicines and preventing counterfeit drugs, a digital drug control system (DDCS) provides a durable solution in the market. Some pharmaceutical companies such as Sanofi, Pfizer, and Amgen launched a combine pilot project with blockchain-based DDCS to inspect and evaluate new drugs [46]. Blockchain has the potential to track the production and location of the drugs and medicines at any time and it also used for improved traceability of fake and falsified drugs [47]. This technology also ensures the security of drug supply [48] and guarantees the quality and standard of supplied drugs to the end-users.

### 3.6. Neuroscience Studies

Neuroscience is also an emerging discipline in the healthcare industry and is being investigated and new neural technologies are seeking new paradigms for controlling data and devices with brain commands without mechanical interactions [49]. These neural devices interpret the patterns of the brain and translate them into useful instructional commands that are further used for controlling the external devices. These devices also monitor the condition of the brain of a person based on their data in the brain. This special task of reading and translating signals of the human brain is solved with neural interface devices that use several sensitive sensors, computing chips, and wireless communication medium. These devices read the electrical signals in the brain that are further transmitted to the controlled equipment that is a device placed on the head of a person. There are complex algorithms and big data analysis behind all this neural process and they are aimed to use blockchain technology to store these brain signals in the neural interface device.

Blockchain technology is expected to use in the form of information technology in several neuroscience applications such as brain augmentation, brain simulation, and brain thinking. Digitizing and storing all the brain data requires a medium to store that data with security and reliability and blockchain technology provides the facility of brain data storage.

## 4. Use Cases of Blockchain

Blockchain technology is being used by healthcare providers for the sake of security and privacy. Several healthcare organizations are using digital technology to manage their health records for providing an efficient, robust and transparent healthcare system. Blockchain provides the best solution to the data security and management in the healthcare system as secure data with cryptographic codes that are complex to be stolen or misused. The decentralized nature of blockchain architecture enables doctors, patients, and other healthcare providers to access the same information and records once saved in the blockchain ledger safely and quickly. There are various healthcare organizations using blockchain in healthcare security, medical records, medical supply chain, and genomic market for keeping their infrastructure incorruptible and transparent.

**Table 1: Use cases of Blockchain in Various Companies**

| Healthcare Domain | Organization | Industry | Description | White Paper |
|---|---|---|---|---|
| | BurstIQ | Big Data, Cybersecurity | Blockchain for Safe and secure management of patient records with HIPAA rules | [50] |
| **Health Security** | MedicalChain | Electronic Health Record, Medical | Maintains the integrity of health records with a single point of truth.<br>protects patient identity | [51] |
| | Factom | IT, Enterprise | creates products for securing digital records<br>Blockchain for securing digital health records | [52] |
| | Guardtime | Cybersecurity, Blockchain | Helps healthcare companies and governments to implement blockchain into their cybersecurity methods | [53] |
| **Medical Records** | Coral Health | Healthcare, IT | Blockchain to accelerate the care process, automate administrative processes and improve health outcomes. | [54] |
| | Patientory | Blockchain, Cybersecurity, Healthcare, | End-to-end encryption ensures that patient data is shared safely and efficiently. | [55] |
| | ROBOMED | Blockchain, Medicine | Combines AI and blockchain to offer patients a single point of care | [56] |
| | SimplyVital Health | AI, Blockchain, Enterprise Software, | Nexus Health platform is an open-source database that allows healthcare providers, on a patient's blockchain, to access pertinent information. | [57] |
| **Medical Supply Chain** | Blockpharma | Blockchain, Pharmaceuticals, Supply Chain | offers a solution to drug traceability and counterfeiting with the help of a blockchain-based SCM system | [58] |
| | TIERION | SaaS, Blockchain | The company uses blockchain timestamps and credentials to maintain proof of ownership throughout a medical supply chain. | [59] |
| | CHRONICLED | Blockchain, Supply Chain Management | Builds blockchain networks that demonstrate chain-of-custody help pharma companies make sure their medicines arrive efficiently, and they enable law enforcement to review any | [60] |

| | | | suspicious activity like drug trafficking | |
|---|---|---|---|---|
| | Centers for Disease Control and Prevention (CDC) | Government Agency, Healthcare, Security | Uses blockchain to monitor diseases in a supply chain-like manner | [61] |
| **Genomic Market** | Nebula Genomic | Biotechnology, Genetics | Using distributed ledger technology to eliminate unnecessary spending and middlemen in the genetic studying process. | [62] |
| | EncrypGen | Blockchain, Data-Sharing | Gene-Chain blockchain-backed platform that facilitates the searching, sharing, storage, buying and selling of genetic information | [63] |
| | Doc.AI | AI, Blockchain, Medical, Software | Uses machine intelligence, like AI, to decentralize medicine on the blockchain | [64] |

## 5. Challenges

Blockchain technology is an emerging technology and is being used in various sectors with its potential benefits and opportunities to revolutionize the digital world. But the technology is still immature and has its challenges that should be considered to solve in the future implementations. Here we have some major challenges discussed here as under:

**Interoperability:** Blockchain has the major issue of interoperability. As interoperability enables multiple users to send and share data and transactions in the network without any intermediary, but sometimes standards are ignored in the building of blockchain to get more freedom in the network and this causes interoperability and communication issues in the blockchain network.

**Uncertainty:** The blockchain concept is still young and cannot be used without certain specifications and surety. At this time a few successful initiatives are utilizing this modern technology. This challenge is imperative to consider for the successful implementation of blockchain in uncertain situations.

**Storage Capacity:** In the healthcare industry, there is a massive amount of medical data, images, documents, and lab results and it requires a significant space for storing all these types of data. Every node in the blockchain network has a copy of all records, this can lead to the shortage of storage capacity of current blockchain technology.

**Cost:** The cost establishment and maintenance of healthcare records using blockchain is unknown to many of the organizations and no one can adopt the technology without knowing the exact cost and expenses.

Some common strengths, weaknesses, opportunities, and threats of blockchain technology with the healthcare industry are depicted in Figure 4 with a SWOT analysis technique.

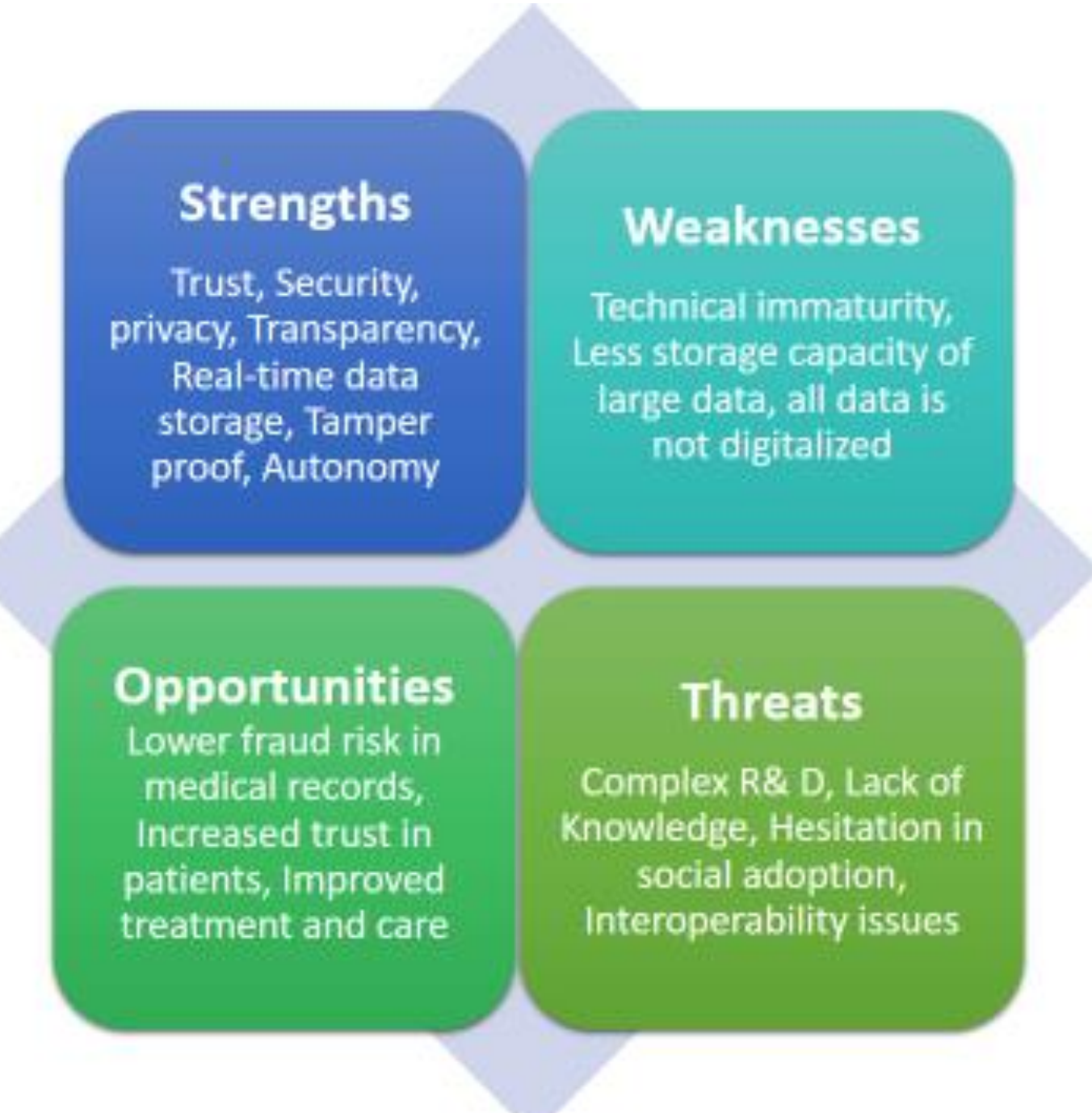

**Figure 4: SWOT Analysis of Blockchain in Healthcare**

## 6. Future Perspectives

Blockchain technology offers several benefits and future perspectives in the healthcare and medicine industry. With the revolutionized healthcare system, blockchain reduces the cost of monitoring, evaluating, and configuring the sensitive medical data without any central authority as a middleman. The future of clinical data can be more secure and reliable where all the records of patients will be available to all the entities involved in the healthcare research and development.

The records stored on the distributed ledger of blockchain will provide a transparent history of patients, so the doctors will not require an honest history of patients form themselves as they have on the ledger without any errors. Patients will also have complete access to their records and they will be free to share their data with various healthcare providers to make better recommendations about their health condition.

## 7. Conclusion

Blockchain technology has potential applications to overcome various challenges faced by the healthcare industry. The strongest potential of blockchain technology is to provide security, integrity, decentralized access to the patients' records, availability of records and authentication of accurate data storage due to its salient features of decentralization, immutability, transparency, and interoperability. The utilization of blockchain in the healthcare industry has many benefits for several individuals including patients, doctors, physicians, healthcare providers, clinical researchers, external healthcare entities, biomedical and neurology experts. They can easily disseminate a large amount of data, share clinical data and results, and can communicate and recommend their patients and other entities involved in the healthcare system with security and privacy. Blockchain integration with clinical research and development will open new research dimensions in the biomedical field. In pharmaceutical medicine and supply chain of drugs and medicines, blockchain will provide safe, secure, and reliable production, storage, and distribution of authentic drugs for better and timely diagnosis and treatment of patients according to their conditions. However, Neurology is still at the experimental stage and less work is done practically with blockchain integration in the brain studies. The blockchain would certainly benefit in the storage of brain data with transparency and immutability as the brain data is more sensitive and need to be handle with care. Consequently, the blockchain technology in healthcare applications will engage entities and individuals to improve their quality of life with privacy and reliability.